\documentclass[twocolumn,showpacs,floatfix,eqsecnum,prb]{revtex4}
\usepackage{float}
\usepackage{graphicx}
\usepackage{amsmath}
\usepackage{bbold}

\DeclareMathOperator{\Tr}{Tr}
\DeclareMathOperator{\Real}{Re}
\DeclareMathOperator{\sgn}{sgn}

\begin{document}
\title{Signatures of Majorana Fermions in Topological Insulator Josephson Junction Devices}

\author{Benjamin J. Wieder, Fan Zhang and C. L. Kane}
\affiliation{Department of Physics and Astronomy, University of Pennsylvania,
Philadelphia, PA 19104}

\begin{abstract}
We study theoretically the electrical current and low-frequency noise for a linear Josephson junction structure on a topological insulator, in which the superconductor forms a closed ring and currents are injected from normal regions inside and outside the ring.  We find that this geometry offers a signature for the presence of gapless 1D Majorana fermion modes that are predicted in the channel when the phase difference $\varphi$, controlled by the magnetic flux through the ring, is $\pi$.   We show that for low temperature the linear conductance {\it jumps} when $\varphi$ passes through $\pi$, accompanied by {\it non-local correlations} between the currents from the inside and outside of the ring.  We compute the dependence of these features on temperature, voltage and linear dimensions, and discuss the implications for experiments.

\end{abstract}

\maketitle

\section{Introduction}
\label{sec:introduction}

There is presently a major effort in condensed matter physics to demonstrate the unique properties of Majorana fermion quasiparticle states associated with topological superconductivity.~\cite{kitaev,readgreen}  A promising approach is to utilize proximity effect devices that combine ordinary superconductors with topological insulators or other strong spin-orbit materials to achieve topological superconductivity.~\cite{fk08,sau10,alicea10,lutchyn10,oreg10}  Recent experiments on semiconductor nanowires coupled to superconductors observed zero-energy tunneling resonances that have been interpreted as Majorana bound states.~\cite{mourik12,deng12,rokhinson12,heiblum12}   The original proposal involved a superconductor coupled to the surface of a topological insulator (TI). It was shown that a vortex in the superconductor is associated with a Majorana bound state, and that a linear Josephson junction (JJ) exhibits gapless 1D Majorana fermions when the phase difference is $\varphi = \pi$.~\cite{fk08}  Supercurrents in TI JJ devices have recently been observed.~\cite{sacepe11,veldhorst12,dgg12}  Unusual behavior, including a smaller than expected critical current normal resistance product, as well as an anomalous Fraunhofer diffraction pattern, has been interpreted as evidence for 1D Majorana fermions along the channel between the superconductors.~\cite{dgg12}  However, the connection between these observations and Majorana fermions is indirect.

A difficulty with critical current measurements is that the predicted Majorana behavior is only manifest when the phase difference $\varphi$ is close to $\pi$.  In order to isolate the properties of the gapless Majorana mode, it is necessary to control the phase.  Moreover, the supercurrent carried by the junction includes contributions from gapped states.  The gapless mode leads to only a weak singularity in the current phase relation at $\varphi = \pi$.  Another possible experiment would be to use a ring geometry and to tunnel into the junction region from another contact.  Similar experiments on 1D SNS junctions revealed the expected collapse of the minigap in the normal region for $\varphi=\pi$.~\cite{esteve08} Such experiments on a TI JJ device could demonstrate the closing of the gap, but they would not distinguish an even and odd number of gapless channels.   The unique feature of the 1D Majorana mode is that as the phase is tuned through $\pi$, the system {\it must} pass through a state where the Majorana mode is transmitted perfectly along the channel - even in the presence of strong disorder.
This leads to a quantized thermal conductance that in principle probes the central charge $c=1/2$ associated with the Majorana mode, but would be difficult to measure.

\begin{figure}
\centering
\includegraphics[width=3.0in]{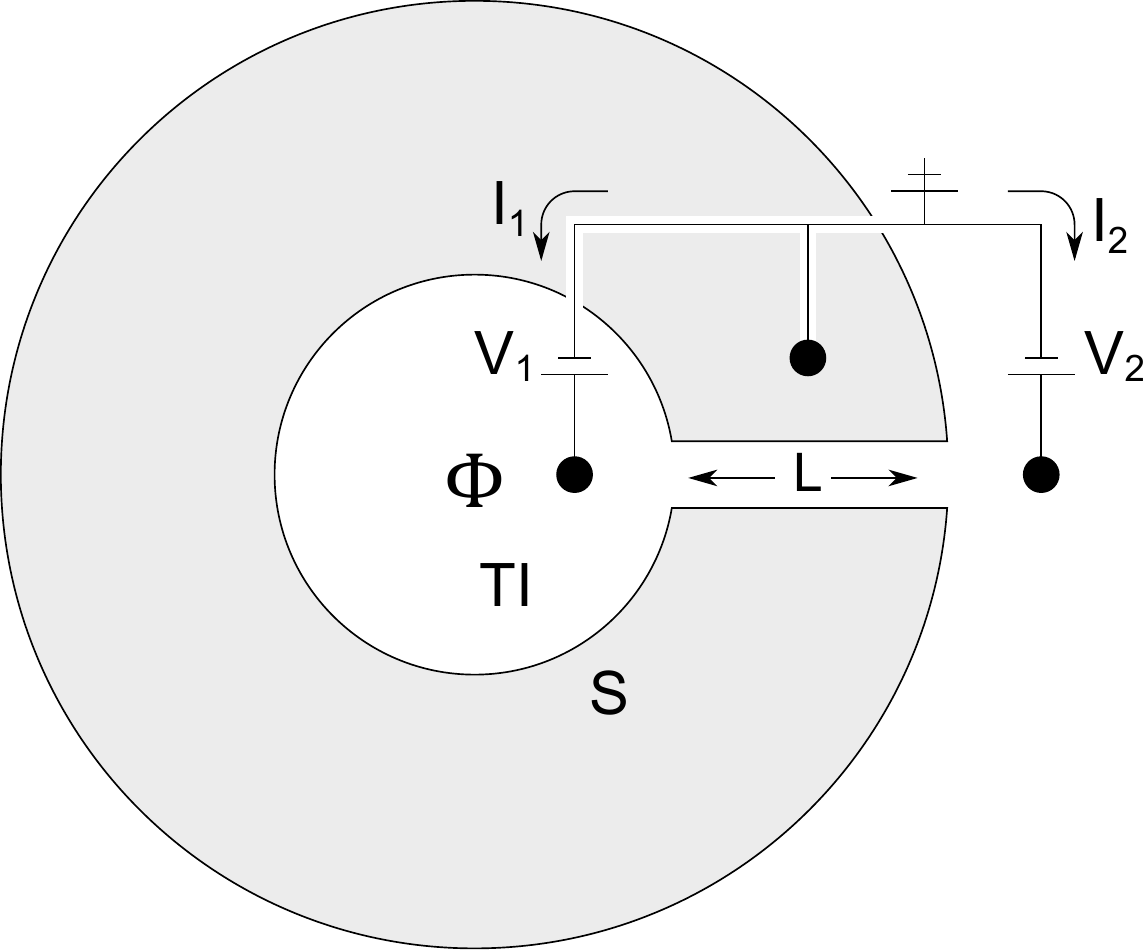}
\caption{(a) Three-terminal ring geometry, in which the phase difference across a linear S-TI-S Josephson junction is controlled by the magnetic flux $\Phi$ through the ring.  Measurement of the current and low-frequency noise provides a signature sensitive to Majorana fermion modes on the junction.  }
\label{Fig1}
\end{figure}
In this paper we show that measurement of current and low-frequency current noise injected from the normal surface states at the {\it ends} of a TI JJ give rise to a unique signature that is sensitive to the perfectly transmitted gapless Majorana mode at $\varphi=\pi$.  Our setup, shown in Fig.~\ref{Fig1}, consists of a superconducting ring on the surface of a topological insulator with a linear channel.  The phase difference across the junction is controlled by the magnetic flux through the ring, $\varphi = 2\pi \Phi/\phi_0$, with $\phi_0 = h/2e$.~\cite{rashbanote}  In addition, we introduce electrical contacts to the normal TI surfaces inside and outside the ring, as well as to the superconductor.   This three-terminal geometry allows the average currents $\langle I^a \rangle$ ($a=1,2$) and the correlations $\langle \delta I^a \delta I^b \rangle$ to be measured as functions of the voltages $V^a$ relative to the grounded superconductor.   We find that for a long junction at low temperature the linear conductance exhibits sharp steps when $\varphi$ passes through odd multiples of $\pi$.  This singular behavior is a direct consequence of the gapless Majorana mode at $\varphi=\pi$.   When the electrical contacts consist of only a single transmitting channel, the magnitude of these steps is $2e^2/h$.  In the more realistic situation where there are multiple active transmitting channels the singular behavior is still present, though the size of the step is reduced.

In addition, we find signatures for the Majorana mode in the diagonal and cross correlations of the current noise.  The cross correlation $\langle \delta I^1 \delta I^2\rangle$ exhibits a narrow peak when $\varphi \sim \pi$ due to the Majorana mode.   In the single-channel limit, the height of the peak in the zero temperature limit is universal.  For many channels, the peak height is suppressed.  The magnitude and sign of the peak, however, is predicted to be related to the size of the steps measured in the average current.   Finally, we predict that the diagonal noise correlation $\langle \delta I^1 \delta I^1\rangle$ also exhibits a sharp peak at $\varphi=\pi$.  Unlike the singular behavior of the average conductance and the cross correlation, however, the size of the peak is {\it not} suppressed when there are many channels, and provides a more robust signature of the Majorana mode.

Our results are related to earlier work concerning resonant transmission through 0D Majorana bound states and proximity effect systems.~\cite{demler07,nilsson08,akhmerov09,fk09,law09,flensberg10,chung11,liu11,strubi11,akhmerov11,wimmer11,buettiker12,beri12,Nagaosa}  In particular,
Law, Lee and Ng~\cite{law09} showed that tunneling into a Majorana bound state at $T=0$ leads to a zero-bias resonant conductance $2e^2/h$ associated with perfect Andreev reflection of a single channel.  In our geometry there are no discrete Majorana zero modes.  However, when the magnetic flux, rounded to the nearest integer multiple of $\phi_0$ is odd, there is effectively a Majorana zero mode inside the ring, but it is strongly coupled to the continuum of states in the surface region coupled to the lead.  In this regard, the linear JJ exhibits a behavior similar to the topological transition in a 1D topological superconductor~\cite{akhmerov11}.  In our case, the transition between the topological and non-topological phase, which occurs at $\varphi=\pi$ is controlled by the magnetic flux.   Noise correlations associated with tunneling into Majorana bound states have also been studied, and the noise correlations for $\varphi=\pi$ resembles the noise correlations that have been studied for coupling to {\it chiral} Majorana fermion modes associated with magnet-superconductor interfaces on the surface of a TI.~\cite{law09,buettiker12}  An important difference between the present work and these earlier works, however, is that in our geometry, the superconducting phase $\varphi$, controlled by the magnetic flux through the ring, provides an accessible knob for controlling the coupling between the counterpropagating chiral Majorana modes.  This makes it possible to tune through the quantum critical point at $\varphi=\pi$ that separates topological and non-topological phases.  An advantage of the present approach is that the singular behavior of the current and noise near $\varphi = \pi$ provides a distinctive signature for the Majorana physics.   

The organization of our paper is as follows.
We first present in section~\ref{sec:modelsystem} the specifics of our model for the Josephson junction, focusing on a simple tunneling problem for the Majorana channel and properly treating the interchannel reflection of the remaining modes in the leads.  Section~\ref{sec:experimentalsignatures} focuses on experimental signatures of the gapless Majorana channel:  first we discuss the ideal case of single-channel leads and then we expand out to the more realistic multi-channel case, extracting the terms in the multi-channel current and noise which show singular behavior as $\varphi$ winds through $\pi$. Finally, we conclude with a discussion of experimental parameters and feasibility.
Additional appendixes~\ref{appendix:averagecurrent} through~\ref{appendix:diagonalnoise} provide complete derivations of the multi-channel observables and their single- and many-channel limits.

While this manuscript was in the final stages of preparation we received a preprint by Diez, et al. that presents an analysis of a related Josephson junction geometry for topological superconductors.~\cite{diez13} 

\section{Model System}
\label{sec:modelsystem}

\begin{figure}
\centering
\includegraphics[width=3.3in]{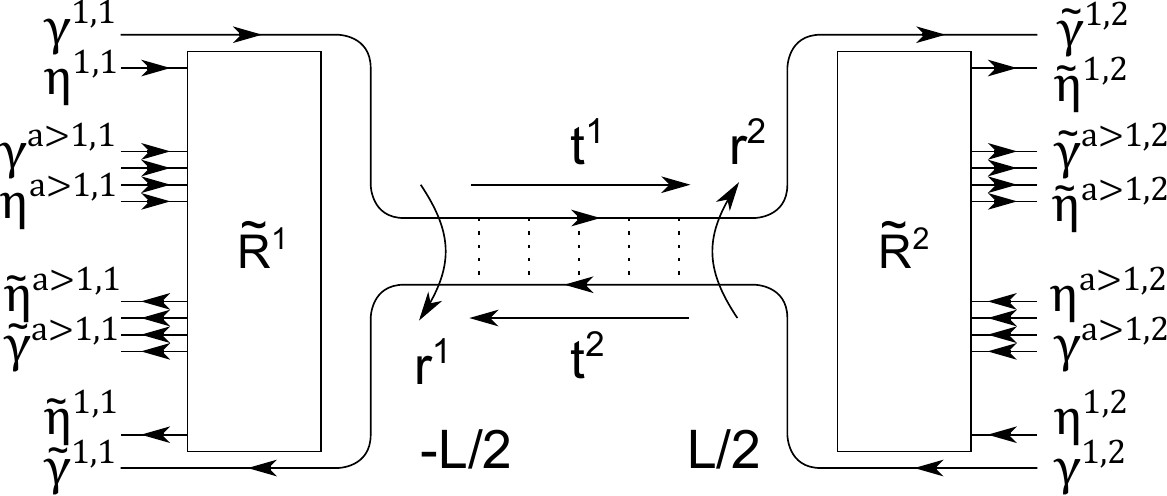}
\caption{ Schematic of the 1D Majorana modes that propagate along the junction for $\varphi\sim\pi$, connecting reservoirs on the inside and outside of the ring.  A mass term $\Delta_0\cos(\varphi/2)$ couples the counterpropagating Majorana modes.   The transmitted mode at $\varphi=\pi$ defines a single Majorana mode in both contacts.  The other Majorana modes in the contacts will be reflected, and characterized by reflection matrices $\tilde R^1$ and $\tilde R^2$.}
\label{Fig1b}
\end{figure}

In Ref.~\onlinecite{fk08} it was shown that a Josephson junction on the surface of a topological insulator exhibits a gapless Majorana mode when the phase difference is equal to $\pi$.  At that point, there is a single, one-dimensional Majorana mode that is transmitted perfectly along the length of the junction.  For $\varphi$ different from $\pi$, a gap opens in that mode.   In the following we consider the low energy limit for phase difference close to $\pi$, so that the transmission across the junction is dominated by the single gapless Majorana mode, which can be described by a simple one-dimensional model.  As indicated in Fig. \ref{Fig1b}, this mode couples to a single Majorana mode in the contacts inside and outside the ring.  In general, the contacts will involve many additional channels, so it is necessary to introduce a general scattering matrix that relates the incident channels and the transmitted channel.

We begin with a discussion of the one-dimensional transmission problem for the singular channel in \ref{sec:1dmodel} and in \ref{sec:nchannelscattering} we introduce the general scattering matrix, from which we compute the current and noise as functions of the phase difference $\varphi$, temperature $T$ and the voltages $V^{1,2}$ on the inner and outer contacts relative to the grounded superconductor.

\subsection{One-Dimensional Model}
\label{sec:1dmodel}

In Ref.~\onlinecite{fk08} a TI JJ was described by modeling the TI surface state by a single massless 2D Dirac fermion coupled to a superconducting pairing potential $\Delta(y) = \Delta_0 \exp (i {\rm sgn}(y) \varphi/2)$.  For $\varphi \sim \pi$ there are quasiparticle states with $E \sim 0$ bound to the interface $y=0$ that are described by a two-band Hamiltonian,
$H = ( \gamma^L , \gamma^R) {\cal H} ( \gamma^L , \gamma^R)^T$, where the one-body Bogoliubov de Gennes Hamiltonian is\cite{similar model}
\begin{equation}
{\cal H} = -i \hbar v_F \sigma^z \partial_x + m(x) \sigma^y\,.
\label{h0}
\end{equation}
Here $\gamma^a(x) = \gamma^{a}(x)^\dagger$ are 1D Majorana fermion operators with $a=L,R$.  In this Majorana basis the one-body Hamiltonian ${\cal H}$ exhibits particle-hole symmetry $\{H,\Xi\}=0$ with $\Xi = K$, complex conjugation.  The mass term is $m = \Delta_0 \cos \varphi/2$.  Importantly, $m$ changes sign when $\varphi$ advances by $2\pi$.  In our ring geometry, this means that ${\rm sgn}(m) = (-1)^\nu$, where $\nu = \Phi/\phi_0$, rounded to the nearest integer.  The mass term violates time-reversal symmetry, expressed by $[{\cal H},\Theta]=0$ with $\Theta = i\sigma^y K$.  For $\varphi = \pi$, $m=0$, there are uncoupled counterpropagating chiral Majorana fermion modes on the 1D interface.  

We should note that the presence of local time-reversal symmetry in the junction at $\varphi=\pi$ obscures the fact that time-reversal symmetry is explicitly broken globally.  Furthermore, the physics of this system requires that time-reversal symmetry be broken generally throughout the system.  To see this, consider that there is only a single pair of counterpropagating Majorana modes in the junction at $\varphi=\pi$.  In a one-dimensional system like this junction, it is not possible to have an odd number of Majorana Kramers pairs without breaking time-reversal symmetry in at least some part of the system.  Consequently, our system corresponds to symmetry Class D and therefore, even at $\varphi=\pi$, it is appropriate to consider time-reversal symmetry globally broken in our setup~\cite{AZsym}.  This classification is more than semantic; the restoration of system-wide time-reversal symmetry at $\varphi=\pi$ would generate a second pair of Majorana modes in the junction which would be subject to additional interactions and could alter experimental signatures~\cite{diez13,fan13}.  Despite the global breaking of time-reversal symmetry, it is worth clarifying that for $\varphi=\pi$, the subsystem of the leads, the junction, and the TI surface linking them {\it does} have local time-reversal symmetry.  This symmetry prevents the transmitting Majorana mode from being backscattered by non-magnetic disorder.  Therefore, one may still analyze moderate disorder in this subsystem by exploiting a time-reversal symmetry, as we later will in \ref{sec:current}.  To summarize more precisely, the global breaking of time-reversal symmetry dictates that the junction hosts just a single counterpropagating Majorana Kramers pair at $\varphi=\pi$, whereas the local preservation of time-reversal symmetry protects the transmission of those paired modes into the leads.      

To model the ends of the junction, we suppose that $\Delta(x,y)$ varies adiabatically as a function of $x$ and smoothly goes to zero in the lead regions with $|x| > L/2$.  In this case each of the 1D Majorana modes in the junction evolves into one of the many propagating channels in the leads.  In the spirit of the Landauer-B\"uttiker approach,~\cite{lb} we focus on this single channel and arrive at a 1D model  described
by Eq.~(\ref{h0}) for the finite length JJ coupled to the leads.  This defines a scattering problem for the chiral Majorana fermion modes incident from the leads.

This scattering problem can be characterized by a $2\times 2$ $S$ matrix,
\begin{equation}
S_E = \left(\begin{array}{cc} r^1_E & t^2_E \\ t^1_E & r^2_E \end{array}\right)\,,
\end{equation}
where $t^{1(2)}_E$ and $r^{1(2)}_E$ describe the amplitudes for transmission and reflection of quasiparticles with energy $E$ incident from the left (right) side. $S_E$ obeys a number of general constraints.  Unitarity requires $|r^{\alpha}_{E}|^2 + |t^{\alpha}_{E}|^2=1$ and $|t^{1}_{E}|^{2}=|t^{2}_{E}|^{2}=|t_{E}|^{2}$.  Particle-hole symmetry requires $S_{-E} = S_E^*$.  The scattering problem is easily solved for the simple model $m(x) = \theta(L/2-|x|) \Delta_0 \cos \varphi/2$.  This model has a mirror symmetry ($x\rightarrow -x$), under which $S_E \rightarrow \sigma^y S_E \sigma^y$, so that
$t^1_E = t^2_E \equiv t_E$ and $r^1_E = -r^2_E \equiv r_E$.   We find
\begin{eqnarray}
t_E = {\hbar v_F\kappa \over{ \hbar v_F\kappa \cosh \kappa L - i E \sinh \kappa L}}\label{scatteringEqn1}\,, \\
r_E = {m \sinh \kappa L \over {\hbar v_F\kappa\cosh \kappa L - i E \sinh \kappa L}} \label{scatteringEqn2}\,,
\end{eqnarray}
where $\kappa = \sqrt{E^2-m^2}/(\hbar v_F)$.  At $E=0$, $S_0$ is real, and is characterized by
\begin{eqnarray}
t_0 &=& 1/\cosh (m/\Delta\epsilon)\,,\\
r_0 &=& \tanh (m/\Delta\epsilon),
\label{r0t0}
\end{eqnarray}
where we have defined $\Delta\epsilon = \hbar v_F/L$.
For $m \gg \Delta\epsilon$, $r_0 = \pm 1$.
Importantly, when $m$ changes sign, $r_0$ changes sign and {\it must} pass through zero, at which point the transmission at $E=0$ is {\it perfect}, {\it i.e.}, $|t_0|=1$.  This property is more general than our specific model.  It is related to the fact the a discrete Majorana zero mode must be present inside the ring for the enclosed flux $\Phi = n\phi_0$ ($\varphi = 2\pi n$) when $n$ is odd,  but is absent when $n$ is even.
In order for the zero mode to appear or disappear, there must be a point where the gap vanishes and the transmission is perfect for $2\pi n < \varphi < 2\pi (n+1)$.  The perfect resonant transmission is thus a specific signature for a gapless 1D Majorana mode on the junction.   However, the transmitted Majorana mode does not carry charge.  We will see in the following that the transmitted Majorana mode leads to a {\it step} in the average current and a peak in the current noise.

\subsection{Current and Noise}
\label{sec:nchannelscattering}

We now develop general formulas for the electrical current and noise in our geometry.  Similar calculations have been performed previously in Refs.~\onlinecite{demler07,nilsson08,akhmerov09,fk09,law09,flensberg10,chung11,liu11,strubi11,akhmerov11,wimmer11,buettiker12,beri12}.  We must combine the transmitted Majorana mode with an additional Majorana mode in each lead as well as properly treat the remaining incident electron channels.  For each channel, the Dirac fermion electron operators may be expressed in terms of a pair of Majorana operators,
\begin{eqnarray}
c^{a,\alpha}_{E} &=& \gamma^{a,\alpha}_{E} + i\eta^{a,\alpha}_{E}\,,\\
\tilde{c}^{a,\alpha}_{E} &=& \tilde{\gamma}^{a,\alpha}_{E} + i\tilde{\eta}^{a,\alpha}_{E}\,,
\label{cae}
\end{eqnarray}
where $c^{a,\alpha}_{E}\,(\tilde{c}^{a,\alpha}_{E})$ describe modes incoming (outgoing) from lead $\alpha$ and channel $a$.  We should note that, in general, a lead $\alpha$ with $N^{\alpha}$ electron channels will have $2N^{\alpha}$ electron and hole channels constrained by particle-hole symmetry.   We are free to define our modes such that at $\varphi=\pi$, $\gamma^{1\alpha}$ is the extension of the perfectly transmitted mode into the leads.  Thus, there is $1$ transmitting Majorana channel and $2N^{\alpha}-1$ reflected Majorana channels with, in general, no additional constraints. 

We can express the relationship between incoming and outgoing Majorana modes in terms of a scattering matrix,
\begin{eqnarray}
\tilde{\gamma}^{a}_{E} = S^{ab}_{E}\gamma^{b}_{E}\,,
\end{eqnarray}
where $a$ and $b$ are now indexes for lead, channel, and Majorana type ($\gamma$ or $\eta$) and noting that $S^{ab}_{E} = S^{ab*}_{-E}$  due to particle-hole symmetry.  $S_E$ has the general structure
\begin{eqnarray}
S_{E} = \left(
\begin{array}{cc}
\mathbb{r}^{1}_{E} & \mathbb{t}^{2}_{E} \\
\mathbb{t}^{1}_{E} & \mathbb{r}^{2}_{E}
  \end{array}\right)\,,
\label{Smatrix}
\end{eqnarray}
which allowing for only a single transmitting Majorana channel becomes:
\begin{eqnarray}
\mathbb{t}^{\alpha}_{E}&=&t^{\alpha}_{E}\,\mathbb{e}^{11}\,, \\
\mathbb{r}^{\alpha}_{E}&=&r^{\alpha}_{E}\,\mathbb{e}^{11} + \mathbb{\tilde{r}}^{\alpha}_{E}\,,\\
\mathbb{\tilde{r}}^{\alpha}_{E} &=& \left(
\begin{array}{cc}
0 & \vec{0}^{T} \\
\vec{0} & \mathbb{\tilde{R}}^{\alpha}_{E}
  \end{array}\right)\,,
\end{eqnarray}
where $r^{\alpha}_{E}$ and $t^{\alpha}_{E}$ are single-channel scattering coefficients, $\mathbb{\tilde{R}}^{\alpha}$ is a $(2N^{\alpha}-1)\times (2N^{\alpha}-1)$ dimensional Majorana reflection matrix representing the remaining channels, and $\mathbb{e}^{ab}_{ij}=\delta_{ai}\delta_{bj}$.  Our plots use for $r^{\alpha}_{E}$ and $t^{\alpha}_{E}$ the model scattering coefficients in Eq.~(\ref{scatteringEqn1}) and (\ref{scatteringEqn2}).  $S$ and all other matrices for our model are $2(N^{1} + N^{2})\times 2(N^{1} + N^{2})$ dimensional.

The operator for the current flowing out of contact $\alpha$ is given by
\begin{eqnarray}
\hat I^{\alpha} &=&  {e v_F\over L} \sum_{E}\sum_{a=1}^{N^{\alpha}} \left( c_{-E}^{a,\alpha\dagger} c_{E}^{a,\alpha} - \tilde c_{-E}^{a,\alpha\dagger} \tilde c_{E}^{a,\alpha} \right) \nonumber\\
&=&  {e v_F\over L} \sum_{E} \left(\gamma_{E}^{\dagger}\Sigma_{y}^{\alpha}\gamma_{E}- \tilde{\gamma}_{E}^{\dagger}\Sigma_{y}^{\alpha}\tilde{\gamma}_{E}\right)\nonumber\\
&=&  {e v_F\over L} \sum_{E} \gamma_{-E}^{a}A^{ab,\alpha}_{E}\gamma_{E}^{b}\,,
\end{eqnarray}
where $A^{\alpha}_{E} = \Sigma_{y}^{\alpha} - S^{\dagger}_{E}\Sigma_{y}^{\alpha}S_{E}$, $\gamma_{E}^{a}$ here is an element of a column vector of Majorana operators in channel $a$, and $\Sigma_{y}^{\alpha} = P^\alpha \otimes \sigma^y$, where $P^\alpha$ is a projector onto the modes in lead $\alpha$ and $\sigma^y$ is the Pauli matrix coupling  $\gamma^{a,\alpha}$ and $\eta^{a,\alpha}$. Omitted details for this calculation are presented in Appendix~\ref{appendix:averagecurrent}.
The average current $I^{\alpha} = \langle\hat{I}^{\alpha}\rangle$ thus reads
\begin{eqnarray}
I^{\alpha} &=& {e v_F\over L} \sum_{E} \langle\gamma^{a}_{-E}\gamma^{b}_{E}\rangle A^{ab,\alpha}_{E} \nonumber\\
&=& {e v_F\over L} \sum_{E} \Tr[Q^{T}_{E}A^{\alpha}_{E}] ,
\label{matrixcurrent}
\end{eqnarray}
in which we have used the following definition
\begin{eqnarray}
Q^{ab}_{E} \equiv \langle\gamma^{a}_{-E}\gamma^{b}_{E}\rangle = \frac{1}{4}\sum_{\beta}\left[f^{+,\beta}_{E}P^{\beta} + f^{-,\beta}_{E}\Sigma_{y}^{\beta}\right]^{ab},
\label{qeq}
\end{eqnarray}
where $f^{\pm,\beta}_E \equiv f^{\beta}(E) \pm [1- f^{\beta}(-E)]$, the sum and difference of the electron and hole Fermi functions.

Similarly, the low-frequency noise power $P^{\alpha\beta}=\int_{-\infty}^{+\infty}dt\big(\langle\hat{I}^{\alpha}(t)\hat{I}^{\beta}(0)\rangle - {I}^{\alpha}{I}^{\beta}\big)$ can be calculated as follows using Wick's theorem.  We find
\begin{eqnarray}
P^{\alpha\beta} &=& \frac{e^{2}v_{F}}{L} \sum_{E,E'} \langle\gamma^{a}_{-E}\gamma^{b}_{E}\gamma^{c}_{-E'}\gamma^{d}_{E'}\rangle A^{ab,\alpha}_{E}A^{cd,\beta}_{E'} \nonumber\\
&=& {2e^2 v_F\over L}\sum_{E}\Tr [A^{\alpha}_{E}Q_{-E}A^{\beta}_{E}Q^{T}_{E}]\,.
\label{matrixnoise}
\end{eqnarray}
The detailed derivation is explained in Appendix~\ref{appendix:crossnoise}.

\section{Experimental Signatures}
\label{sec:experimentalsignatures}

Here we provide a description of the current and noise observables which characterize the 1D gapless Majorana channel in the JJ.  We begin with the average current and conductance at each lead, finding that they are independent of the applied voltage at the other lead and display sharp steps in the low temperature and small voltage limit.  We then consider the noise power across the leads and at the same lead.  The cross noise signal contains only one term which exhibits a peak at $\varphi\sim\pi$, though is suppressed by interchannel scattering.  The diagonal noise displays a more complicated signal but exhibits a peak that persists even in the large $N^{1,2}$ limit.

\subsection{Average Current}
\label{sec:current}

We will begin with a calculation of the average electric current in the limit that there is only a single channel in the electrical contacts.  In principle, this could arise if there was a quantum point contact separating the leads from the surface states.  Even away from this limit, however, the simplicity of the result will aid the understanding of the more general results, which we present in the following section.

\subsubsection{Single-Channel Limit}
\label{sec:1channelcurrent}

\begin{figure}
\scalebox{0.65}{\includegraphics*{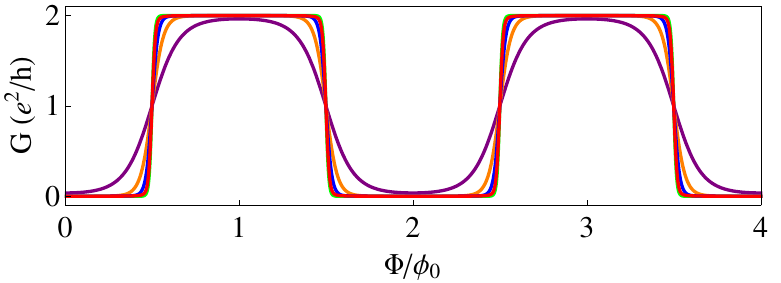}}
\caption{Conductance as a function of $\varphi$ in the single-channel limit. The green curves represent the ideal case for zero temperature and zero voltage in each lead with $\Delta\epsilon =0.05 \Delta_0$. The purple, orange, blue, and red curves respectively represent the cases for $\Delta\epsilon/\Delta_0=0.5$, $0.2$, $0.1$, and $0.05$ with $T=0.005\Delta_0$ and $eV_1=eV_2=eV=0.05\Delta_0$.}
\label{Fig2}
\end{figure}

In the single-channel limit, there is only one additional Majorana mode $\eta^{1,\alpha}$ in each lead, so that ${\mathbb{\tilde{R}}}^\alpha = \pm 1$.  
Using Eq.~(\ref{matrixcurrent}) and~(\ref{matrixnoise}) and setting $\sum_{E}=L/(2\pi\hbar v_F)\int_{-\infty}^{+\infty} dE$, we obtain
\begin{equation}
I^\alpha = {e\over {2h}}\int_{-\infty}^\infty dE (1- r_E^\alpha) f_-^\alpha(E).
\end{equation}
In Fig. \ref{Fig2} we plot  the conductance $G = I/V$ as a function of phase $\varphi$ predicted by Eq.~(\ref{Iav}) for several values of $\Delta\epsilon$.  $G(\varphi)$ exhibits sharp steps at $\varphi = 2\pi (n+1/2)$, provided $\Delta\epsilon \ll \Delta_0$.  This requirement is equivalent to having the coherence length $\xi \sim \hbar v_F/\Delta_0 \ll L$, so that quasiparticle tunneling across the superconductor is suppressed for $\varphi \sim 0$, leading to perfect normal or Andreev reflection, $r(\varphi\sim 0) = \pm 1$.   In this limit the width of the step is determined by the maximum of $\Delta\epsilon$ and $T$.
For $T \ll \Delta\epsilon \ll \Delta_0$ the linear conductance is simply,
\begin{equation}
G(\varphi) = (1- r_0)e^2/h\,
\label{gvarphi}
\end{equation}
where from Eq.~(\ref{r0t0}) $r_0 = \tanh[(\Delta_0/\Delta\epsilon) \cos\varphi/2]$.
$r_0$ switches between $-1$ and $+1$ over a range $\delta\varphi \sim \Delta\epsilon/\Delta_0 \ll 1$, so that over that range $G$ exhibits a step
\begin{equation}
\Delta G = 2 e^2/h.
\end{equation}
In the following section we will show that when there are additional channels, the step is still present, but its magnitude is suppressed.

Eq.~(\ref{gvarphi}) can be understood in terms of the four elementary scattering processes for a particle at the Fermi energy incident from one of the leads.  The probability for reflection as an electron (or hole) is $|1 + (-) r_0|^2/4$, while the probabilities for transmission as an electron or as a hole are both $(t_0)^2/4$.
For $\Delta\epsilon \ll T \ll \Delta_0$, $G(\varphi) =  \tilde G( m_\varphi/T)e^2/h$, with  $\tilde G(X) = 1-X \int_0^1 dz \sqrt{1-z^2}/\cosh^2(X z)$.

\subsubsection{General N-Channel Current}
\label{sec:Nchannelcurrent}

In the general case of many electron channels in each lead we find
\begin{eqnarray}
I^{\alpha} = \frac{e}{4h}\int_{-\infty}^{+\infty} dE\ f^{-,\alpha}_{E}\bigg\{2N^{\alpha} - 2\Real [r^{\alpha*}_{E}\mathbb{\tilde{r}}^{\alpha}_{22,E}] \nonumber \\
-\Tr [\Sigma_{y}\mathbb{\tilde{r}}^{\alpha,\dagger}_{E}\Sigma_{y}\mathbb{\tilde{r}}^{\alpha}_{E}]\bigg\}.
\label{Iav}
\end{eqnarray}
This expression contains terms that do not depend on the scattering of the mode that is perfectly transmitted at $\varphi = \pi$.  In general, these terms will depend on  $\varphi$, but they will {\it not} exhibit the singular $\varphi$ dependence associated with the critical mode.  Thus, we extract the singular terms, which we denote by $I^\alpha_\gamma$, that depend on $r^\alpha$ and $t^\alpha$.
\begin{equation}
I^{\alpha}_{\gamma} = -\frac{e}{2h}\int_{-\infty}^{+\infty} dE\ f^{-,\alpha}_{E}\bigg\{\Real [r^{\alpha*}_{E}\mathbb{\tilde{r}}^{\alpha}_{22,E}]\bigg\}\,.
\label{MajoranaNcurrent}
\end{equation}
In the limit $T, V^\alpha \ll \Delta\epsilon$, we have $S_{E}\approx S_{0}$.  Since $S_{E}=S^{*}_{-E}$, $S_{0}$ is a real, orthogonal matrix.  The conductance jump at $\varphi \sim \pi$ due to the step in $r_0$ then becomes
\begin{equation}
\Delta G^{\alpha}=\frac{2e^{2}}{h}\mathbb{\tilde{r}}^{\alpha}_{22,0}\,.
\label{gJump}
\end{equation}

In general, $\mathbb{\tilde{r}}^{\alpha}_{22,0}$ will depend on the details of the interface between the Josephson junction and the electrical contacts, and can vary in both sign and magnitude.  We will not attempt to compute it in detail here.  Rather, we will note that the scattering of our system lies somewhere between the limits of a disordered, many-channel quantum point contact and that of a diffusive, quasi-1D conductor.  In the limit where the TI surface between the leads and the junction is extremely clean, we can consider, as is done in Ref.~\onlinecite{diez13}, that $\mathbb{\tilde{r}}^{\alpha}_{22,0}$ is an element of a $(2N^\alpha -1) \times (2N^\alpha-1)$ orthogonal matrix which is, in general, unconstrained by time-reversal or spatial symmetries.  Under the assumption that all such matrices are equally likely, the typical value will be $\mathbb{\tilde{r}}^{\alpha}_{22,0} = O(1/\sqrt{N^\alpha})$.  Conversely, as is discussed in Ref.~\onlinecite{imry86}, we can consider that in the limit that the TI surface linking the lead and the junction has moderate, non-magnetic disorder and has a comparable to that of the elastic mean free path, only a limited number of the electron channels on the TI surface will actively carry current.  In this case, only the active channels which penetrate the disorder will be subject to interchannel scattering and the typical value of $\mathbb{\tilde{r}}^{\alpha}_{22,0}$ will be increased to $O(1/\sqrt{N^{\alpha}_{open}})$ where $N^{\alpha}_{open}$ is on the order of the TI surface conductance in units of $e^{2}/h$.  Since the derivation of the experimental signatures is unchanged between the two limits, we will simplify our notation and redundantly label $N^{\alpha}_{open} \equiv N^{\alpha}$ such that for all intermediate cases 

\begin{equation}
\mathbb{\tilde{r}}^{\alpha}_{22,0} = O(1/\sqrt{N^\alpha}).
\end{equation}

Thus, though the magnitude is suppressed, the conductance still exhibits sharp jumps in the limit $T \ll \delta E \ll \Delta_0$.  Importantly, the conductances $G^1$ and $G^2$ from the inside and outside leads should exhibit jumps at the same magnetic flux.  If observed, these conductance jumps represent a clear signal of a quantum phase transition in the system, corresponding to the insertion or removal of a delocalized Majorana mode in the superconducting ring.  The magnitudes and signs of these jumps are each characterized by  $\mathbb{\tilde{r}}^{\alpha}_{22,0}$.  We will see that the same parameters characterizes the signature in the noise correlations and that the cross noise correlation peaks occur at the same flux as the conductance jumps.  

\subsection{Noise Power}

Next we calculate the current-current correlations which contribute to zero-frequency noise power $P^{\alpha\beta}$.  We find that the cross correlation $P^{12}$ is characterized be a peak at $\varphi\sim\pi$ due to a single term which corresponds to Majorana transmission across a 1D gapless channel and if detected at half integer multiples of $\phi_{0}$ gives an unambiguous signature of a quantum phase transition and the insertion of a delocalized Majorana mode into the ring.   For a single channel, the peak has a universal height, while, for many channels it is suppressed by the same factors that led to the suppression of the conductance steps.  The diagonal noise $P^{\alpha\alpha}$ will be presented in its single-channel and many-channel limits with relevant behavior highlighted.  Unlike the cross noise, the diagonal noise contains singular terms which remain as $N^{1,2}\rightarrow\infty$.

\subsubsection{Single-Channel Limit}
\label{sec:1channelnoise}

Taking Eq.~(\ref{matrixnoise}) in the previously-discussed single-channel limit we find
\begin{eqnarray}
P^{12} &=& \frac{e^{2}}{4h}\int_{-\infty}^{+\infty} dE f^{-,1}_{E}f^{-,2}_{E}t^{1}_{E}t^{2}_{E}
\,,\label{p12}\\
P^{11} &=& {e^2\over{4h}} \int_{-\infty}^{+\infty} dE  [|1-r^{1}_{E}|^2 f^{+,1}_{-E}f^{+,1}_{E}  \nonumber \\
&-& (1-r^{1}_E)^2 f^{-,1}_{E}f^{-,1}_{E} + |t_E|^2 f^{+,1}_{-E}f^{+,2}_{E}]\,.\label{p11}
\end{eqnarray}
$P^{21}$ and $P^{22}$ follow from interchanging superscripts $1 \leftrightarrow 2$ in Eq.~(\ref{p12}) and~(\ref{p11}), respectively.

\begin{figure}
\scalebox{0.65}{\includegraphics*{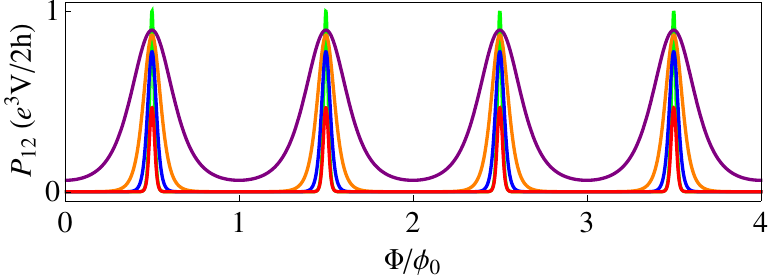}}
\caption{Cross noise power $P^{12}$ as a function of $\varphi$ in the single-channel limit. The green curves represent the ideal case for zero temperature and zero voltage in each lead with $\Delta\epsilon =0.05 \Delta_0$. The purple, orange, blue, and red curves respectively represent the cases for $\Delta\epsilon/\Delta_0=0.5$, $0.2$, $0.1$, and $0.05$ with $T=0.005\Delta_0$ and $eV_1=eV_2=eV=0.05\Delta_0$.}
\label{Fig3}
\end{figure}

Fig.~\ref{Fig3} shows the non-local noise correlation $P^{12}(\varphi)$ evaluated for $e V_1 = e V_2 = 10 T$, for representative values of $\Delta\epsilon$.  For $\Delta\epsilon \ll \Delta_0$, $P^{12}(\varphi)$ exhibits a peak near $\varphi = 2\pi(n+1/2)$.
For $\Delta\epsilon < T$ the peak height is suppressed by a factor $\sim \exp(- \pi T/\Delta\epsilon)$.  Observation of the peak in the noise correlations requires $eV, T \lesssim \Delta\epsilon$.

\begin{figure}
\centering
\includegraphics[scale=0.69]{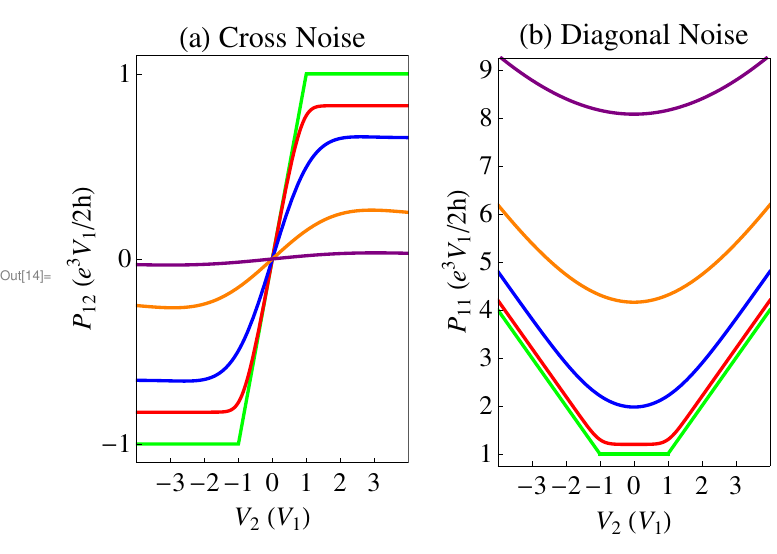}
\caption{Cross noise and diagonal noise at $\varphi=\pi$ in the single-channel limit. The green curves represent the ideal case for zero temperature and zero voltage in each lead. The purple, orange, blue, and red curves respectively represent the cases for $T/\Delta_0=0.1$, $0.05$, $0.02$, and $0.005$ with $eV_1=0.05 \Delta_0$ and $\Delta\epsilon=0.1\Delta_0$.}
\label{Fig4}
\end{figure}

In the limit $eV,T \ll \Delta\epsilon$ Eq.~(\ref{p12}) and~(\ref{p11}) reduce to
\begin{eqnarray}
\tilde P^{11} &=& \frac{e^2}{2h}T\bigg\{(t_{0})^{2}\bigg[F\left(\frac{eV^{+}}{2T}\right) + F\left(\frac{eV^{-}}{2T}\right)\bigg] \nonumber \\
&+& (1-r_{0})^2\bigg\} \,,\\
\tilde P^{12} &=& \frac{e^{2}}{2h}T(t_{0})^{2} \bigg[F\left(\frac{eV^{+}}{2T}\right) - F\left(\frac{eV^{-}}{2T}\right)\bigg]\nonumber \\ \label{crosscoth}\ 
\end{eqnarray}
where $F(X)=X\coth(X)$, $V^{\pm}=V^{1}\pm V^{2}$, and we've made the assumption that $t^{1}_{0}=t^{2}_{0}=t_{0}$ and $r^{1}_{0}=-r^{2}_{0}=r_{0}$.  This assumption is generally valid as most experimental systems in this geometry will be adiabatically connected to our model system in this low-energy limit.  This leads to a striking behavior in the zero-temperature limit.
For $T\ll V_1, V_2 \ll \Delta\epsilon$, we find that
\begin{eqnarray}
P^{11} &=& \frac{e^3}{2h}(t_0)^2 |V_{\rm max}|\,,\\ 
P^{12} &=& \frac{e^3}{2h}(t_{0})^{2} V_{\rm min} {\rm sgn}(V_{\rm max})\,,  \label{p12vmax}
\end{eqnarray}
where $V_{\rm max}  (V_{\rm min})$ is the voltage $V^{\alpha=1,2}$ that has the largest (smallest) absolute value.
Obviously the diagonal (off diagonal) noise correlations are sensitive to the maximum (minimum) of the voltages of the leads relative to the superconductor.    In Fig.~\ref{Fig4} we plot the noise correlations at the peak $m=0$ as a function of $V_2$ for fixed $V_1$ and representative temperatures.  The fluctuation in the total current $I^+ = I^1+I^2$ flowing into the superconductor is given by $P^+ = \langle\delta I^{+}\delta I^{+}\rangle$ which has the following simple form:
\begin{eqnarray}
P^+ = \frac{e^3}{h}t_0^2 |V_1+V_2|\,.
\end{eqnarray}
The current flowing across the junction, $I^- = I^1-I^2$ has noise $P^- = \langle\delta I^{-}\delta I^{-}\rangle$ which goes as
\begin{eqnarray}
P^{-} = \frac{e^3}{h}t_0^2 |V_1-V_2|\,.
\end{eqnarray}

The features in $G(\varphi)$ and $P^{ab}(\varphi)$ near $\varphi=\pi$, which are predicted to occur over a width $\delta\varphi \sim {\rm max}(\Delta\epsilon, eV, T)/\Delta_0$ constitute a signature for the Majorana fermion modes associated with the Josephson junction.  They are present because over the range $\delta\varphi$, a Majorana zero mode is transferred from one end of the junction to the other.

\subsubsection{N-Channel Generalization of the Noise Power}
\label{sec:generalNchannel}

When there are multiple channels the cross correlation is
\begin{eqnarray}
P^{12}&=&P^{12}_{\gamma}\nonumber\\
&=&-\frac{e^{2}}{4h}\int_{-\infty}^{+\infty}dE f^{-,1}_{E}f^{-,2}_{E}t^{1}_{E}t^{2}_{E}\mathbb{\tilde{r}}^{1*}_{22,E}\mathbb{\tilde{r}}^{2*}_{22,E}\nonumber \\
\label{Ncrossnoise}
\end{eqnarray}
In the limit, $eV,T \ll \Delta\epsilon$, the cross noise still maintains the form of Eqn. (\ref{crosscoth}) but is suppressed:
\begin{equation}
\tilde P^{12} = \frac{e^{2}}{2h}T(t_{0})^{2}\bigg[F\left(\frac{eV^{+}}{2T}\right) - F\left(\frac{eV^{-}}{2T}\right)\bigg]\mathbb{\tilde{r}}^{1}_{22,0}\mathbb{\tilde{r}}^{2}_{22,0} .
\end{equation}
For $T\ll V_1, V_2 \ll \Delta\epsilon$, the Fermi functions become step functions and the cross noise becomes 
\begin{equation}
P^{12}(\varphi)=-\frac{e^{3}V_{min}}{2h}\,\sgn(V_{\rm max})\,t_{0}^{2}(\varphi)\,\mathbb{\tilde{r}}^{1}_{22,0}\,\mathbb{\tilde{r}}^{2}_{22,0}\,.
\end{equation}
which still maintains the same $V_{\rm min}$, $V_{\rm max}$ dependence as \ref{p12vmax}.  The height of this peak as a function of $\varphi$ is related to the conductance jumps in Eq.~(\ref{gJump}) by the scattering parameters $\mathbb{\tilde{r}}^{1,2}_{22,0}$ which can be independently measured using the average current at each lead.  Thus, for many channels we expect the cross noise to be suppressed by a factor of order $(1/\sqrt{N^{1}N^{2}})$.

The diagonal noise signal $P^{11}$ has a more complicated dependence on $\varphi$ as well as on the elements of $\tilde{\mathbb{r}}^{1}_{E}$ and is fully derived in Appendix~\ref{appendix:diagonalnoise}.   However, we find that in the large $N^{1,2}$ limit, in which the $O(1/\sqrt{N^\alpha})$ terms are completely suppressed, there remains a peak in the noise that gives a robust signature for the transmitted Majorana mode.   In particular, we find
\begin{eqnarray}
P^{11}_{many} &=& {e^2\over 4h}\int_{-\infty}^{+\infty} dE\ f^{+,1}_{-E}f^{+,1}_{E}\big(2N^{1} -|t_{E}|^2\big) \nonumber \\
&+&f^{+,1}_{-E}f^{+,2}_{E}|t_{E}|^{2}.
\end{eqnarray}
In the $T,V \ll \Delta\epsilon$ limit the singular piece of this becomes
\begin{eqnarray}
P^{11}_{many,\gamma} = \frac{e^{2}}{2h}(t_{0})^{2}T\bigg\{F\left(\frac{eV^{+}}{2T}\right) + F\left(\frac{eV^{-}}{2T}\right) \nonumber \\
-F\left(\frac{eV^{1}}{T}\right) -1\bigg\}.\,\,\,
\end{eqnarray}
The voltage dependence of $P^{11}_{many,\gamma}$ is further clarified in the $T\ll V_1, V_2 \ll \Delta\epsilon$ limit:
\begin{equation}
P^{11}_{many,\gamma} = \frac{e^{3}}{2h}(t_{0})^{2}\bigg[|V_{\rm max}| - |V^{1}|\bigg]
\end{equation}
which vanishes if $|V^{1}|\geq|V^{2}|$.  This behavior is quite distinctive and gives a clear indicator of a single gapless Majorana channel.  There will be a peak in the current in lead 1 due to an applied voltage in lead 2, but no peak if a voltage is only applied to lead 1.
In all of these cases, the diagonal noise still contains singular pieces in the large $N^{1,2}$ limit and therefore provides, of all the quantities discussed in this paper, perhaps the most robust signature of gapless Majorana modes.

\section{Conclusion}
\label{sec:conclusion}

In this paper we have computed the electrical current and noise for a Josephson junction structure on the surface of a topological insulator that allows a clear signature of the gapless Majorana mode, predicted at phase difference $\varphi = \pi$.   We predict that the average current exhibits sharp steps as a function of phase difference for a long junction at low temperature and voltage.  The diagonal and off diagonal noise correlations exhibit peaks at $\varphi = \pi$.  The amplitudes of the singular steps and peaks are predicted to be universal in the case where the electrical contacts couple via a single channel.  For $N$ open electron channels, the singularities remain finite, but the current steps are reduced by $1/\sqrt{N}$, while the cross noise correlation is suppressed by $1/N$.  The diagonal noise includes a peak that is not suppressed for large $N$.

We now briefly discuss some relevant issues for experimentally implementing our proposal.  The number of channels of the leads is an important parameter for determining the lower bound on the size of the singular contributions.  For a ring geometry, as in Fig. \ref{Fig1}, this can be roughly estimated as $N^\alpha \sim k_F R^{\alpha}$, where $R^{\alpha}$ is the radius of the inner or outer edge of the ring.   To minimize this, it is clearly desirable to control the Fermi energy of the topological insulator surface states, such that the Fermi energy is close to the Dirac point.  In this case, $N^\alpha \sim E_F R^{\alpha}/\hbar v_F$, where $v_F$ is the velocity of the surface states.  For Bi$_2$Se$_3$, $\hbar v_F \sim 0.3$eV nm, so for $R \sim 1 \mu$ and $E_F \sim 30$meV, $N \sim 10$ channels~\cite{carrierdensity}.   

Additionally, if there is disorder at the interface between the leads and the junction, the number of active electron channels will be decreased and $N$ will then go instead as the conductance of the TI surface in units of $e^{2}/h$.  Typical values for TI surface conductance in these devices range from $N\sim 20$ to $N\sim 200$ depending on sample purity and efforts to tune the Fermi energy~\cite{steinberg2010}.  Also, in the presence of disorder, the noncritical part of the conductance will be dependent on effects, such as enhanced reflectionless tunneling or weak localization, which depend on the magnitude of the applied field~\cite{beenakker93}.  It is desirable to minimize these aperiodic contributions across the addition of a single $\pi$-flux by decreasing the amount of field required to insert one flux quantum.  To that end, one should make the cross-sectional area of the ring as big as possible. Equivalently, one should maximize $H_{c}A_{ring}/\phi_{0}$ where $H_{c}$ is the critical field of the superconductor.  

A final key parameter in our theory is the level spacing $\Delta\epsilon = \hbar v_F/L$.  To observe sharp features in the current and noise at $\varphi=\pi$ we require $\Delta\epsilon \ll \Delta_0$, so that $L$ is larger than the coherence length $\xi =\hbar v_F/\Delta$.   This ensures that for $\varphi\ne\pi$ the transmission of quasiparticles across the superconductor is exponentially suppressed.   Since the noise peak is suppressed for $\Delta\epsilon < T$, observation of non-local noise correlations requires that $L$ not be too large.  There is ample room to satisfy these constraints experimentally.
For example, in Ref.~\onlinecite{dgg12} devices with Ti/Al electrodes ($\Delta_0 \sim 150 \mu$eV) were studied.  While $v_F$ is not known exactly for these devices, an upper bound is the velocity characterizing the Dirac surface states of Bi$_2$Se$_3$, $\hbar v_F \sim 0.3$eV nm.  This leads to $\xi \lesssim 1.8 \mu$m.   For longer junctions, in which $L > L_T = \hbar v_F/T$ or the inelastic length $L_{\rm in}$, the noise correlations will be suppressed, but the step in the conductance as well as the diagonal noise peak remain robust, provided the flux through the ring (and hence the phase $\varphi$) can be controlled with the applied magnetic field.   For this, it is desirable to minimize the self-inductance ${\cal L}$ of the ring, so that the Josephson energy, $E_J\cos 2\pi \Phi/\phi_0$, which tends to quantize the flux, is dominated by $\Phi^2/2{\cal L}$.
In Ref.~\onlinecite{dgg12}, devices with $L \sim 1 \mu$m had critical current $I_c \sim 1 \mu$A.  Using $E_J = \hbar I_c/e$ and ${\cal L} \sim 4\pi R \log R/L$ (for a ring of radius $R$ and thickness $L$), we find that this condition is satisfied for $R \lesssim 100 \mu$m.

\acknowledgments

It is a pleasure to thank David Goldhaber-Gordon for useful discussions.  This work was supported by NSF grant DMR 0906175 and DARPA grant SPAWAR N66001-11-1-4110, and was partially supported by a Simons Investigator award from the Simons Foundation to Charles Kane.

\begin{appendix}

\section{Average Current Calculation}
\label{appendix:averagecurrent}

For the purpose of clarity, we will work out in detail our derivation of the $N$-channel current and noise, beginning with our choice of a Majorana basis and working in this appendix up to the average current and its single-channel and many-channel limits.  
Appendixes~\ref{appendix:crossnoise} and~\ref{appendix:diagonalnoise} will expand upon this work up to the general expression for zero-frequency noise power and its forms in the interchannel and same-channel cases.

We begin with the choice of constructing Majorana operators
\begin{eqnarray}
\gamma^{a}_{E} &=& \frac{1}{2}\big(c^{a\dagger}_{E} + c^{a}_{-E}\big)\,, \\
\eta^{a}_{E} &=& \frac{i}{2}\big(c^{a\dagger}_{E} - c^{a}_{-E}\big)\,,
\end{eqnarray}
where $a$ represents a given channel and these linear combinations have been chosen such that these new operators are $+1$ eigenstates of the particle-hole operator $\Xi=\tau_{x}K$ in the electron-hole basis of
\begin{equation}
\psi_{E} = \left(
  \begin{array}{c}
  c_{E} \\
  c^{\dagger}_{-E}
  \end{array}\right)\,.
\end{equation}
These operators obey the additional property that $\gamma^{\dagger}_{-E}=\gamma_{E}$, such at $E=0$ they obey the Majorana relation $\gamma^{\dagger}_{0}=\gamma_{0}$.  We note that since our new operators are just linear combinations of electron creation and annihilation operators, they still have canonical anticommutation relations $\{\gamma^{a}_{E},\gamma^{b}_{E'}\}=\delta^{ab}\delta_{E,E'}$ and still obey Wick's theorem when calculating higher order correlation functions.  The two kinds of Majorana have valid contractions with themselves and between species, leading to four correlators that will be of use:
\begin{gather}
\langle\gamma^{a}_{-E}\gamma^{b}_{E'}\rangle = \langle\eta^{a}_{-E}\eta^{b}_{E'}\rangle = \frac{1}{4}f_E^{+}\delta^{ab}\delta_{E,E'}\,, \\
\langle\gamma^{a}_{-E}\eta^{b}_{E'}\rangle = -\langle\eta^{a}_{-E}\gamma^{b}_{E'}\rangle = \frac{i}{4}f_E^{-}\delta^{ab}\delta_{E,E'}\,,
\label{Majoranacorrelator}
\end{gather}
where $f^{\pm}_E \equiv f(E) \pm [1- f(-E)]$ and comes from substituting the definitions of our Majorana operators into the above correlators and noting that $\langle c^{a\dagger}_{E}c^{b}_{E'}\rangle = f_{E}\delta^{ab}\delta_{E,E'}$ and $\langle c^{a}_{-E}c^{b\dagger}_{-E'}\rangle = (1-f_{-E})\delta^{ab}\delta_{E,E'}$, the Fermi distributions for electrons and holes respectively.
From here, we can introduce the current operator
\begin{equation}
\hat{I}^{\alpha} = \frac{ev_{F}}{L}\sum_{E}\psi^{a\dagger}_{-E}\Sigma_{z}^{\alpha,ab}\psi^{b}_{E} - \tilde{\psi}^{a\dagger}_{-E}\Sigma_{z}^{\alpha,ab}\tilde{\psi}^{b}_{E}\,.
\end{equation}
$\Sigma^{\alpha}_{z}$ is the matrix for charge-weighted momentum through lead $\alpha$ in the electron-hole basis and we have expanded $\psi_{E}$ to be $2(N^{1}+N^{2})$ dimensional for our two-lead geometry where $N^{\alpha}$ is the number of channels in lead $\alpha$.  The $-E$ for $\psi^{\dagger}$ comes from the delta function in energy that we get by time-averaging and summing over individual operator energies.  $\Sigma^{\alpha}_{z} = P^{\alpha}\Sigma_{z}$ where $P^{\alpha}$ is the projection matrix into the subspace of lead $\alpha$ and $\Sigma_{z} = \mathbb{1}(N^{1}+N^{2})\otimes\sigma_{z}$.  Rotating this into our Majorana basis
\begin{equation}
\hat{I}^{\alpha} = \frac{ev_{F}}{L}\sum_{E}\gamma^{a}_{-E}\Sigma_{y}^{\alpha,ab}\gamma^{b}_{E} - \tilde{\gamma}^{a}_{-E}\Sigma_{y}^{\alpha,ab}\tilde{\gamma}^{b}_{E}\,,
\end{equation}
where we should note that $\gamma^{a}_{E}$ is a Majorana of type $\gamma$ ($\eta$) for odd (even) $a$.  There will always be an even number of modes since we have artificially doubled the electron channels in each lead as to handle both normal and Andreev reflections off the superconductor.  
The alternating pattern of entries in $\gamma_{E}$ allows us to summarize Eq.~(\ref{Majoranacorrelator}) as follows
\begin{gather}
Q^{ab}_{E} \equiv \langle\gamma^{a}_{-E}\gamma^{b}_{E}\rangle = \sum_{\beta=1,2}Q^{\beta,ab}_{E}\,,\\
Q^{\beta, ab}_{E} = \frac{1}{4}\left[f^{+,\beta}_{E}P^{\beta} + f^{-,\beta}_{E}\Sigma_{y}^{\beta}\right]^{ab}\,.
\end{gather}

The outgoing operators $\tilde{\gamma}^{a}_{E}=S^{ab}_{E}\gamma^{b}_{E}$ where $S_{E}$ is a $2(N^{1}+N^{2}) \times 2(N^{1}+N^{2})$ dimensional scattering matrix that obeys the property $S_{E}=S^{*}_{-E}$, which can be derived by Hermitian conjugating $\tilde{\gamma}_{E}$ and using $\gamma^{\dagger}_{E} = \gamma_{-E}$, and is a consequence of particle-hole symmetry.  This allows us to write the current operator in a much more compact form
\begin{gather}
\hat{I}^{\alpha} = \frac{ev_{F}}{L}\sum_{E}\gamma^{a}_{-E}A^{\alpha,ab}_{E}\gamma^{b}_{E}\,,\\
A^{\alpha}_{E} = \Sigma_{y}^{\alpha} - S^{\dagger}_{E}\Sigma_{y}^{\alpha}S_{E}\,.
\end{gather}

With all of these definitions in place, we can finally begin to take the expectation value of $\hat{I}^{1}$:
\begin{eqnarray}
I^{1} =\langle\hat{I}^{1}\rangle &=& {e v_F\over L} \sum_{E} \langle\gamma^{a}_{-E}\gamma^{b}_{E}\rangle A^{1,ab}_{E} \nonumber \\
&=& {e v_F\over L} \sum_{E} Q^{ab}_{E}A^{1,ab}_{E} \nonumber \\
&=& {e v_F\over L} \sum_{E}\Tr [Q^{T}_{E}A^{1}_{E}]\,,
\end{eqnarray}
which gives us a general average current of
\begin{eqnarray}
I^{1} = \frac{ev_{F}}{4L}\sum_{E}\big\{f^{-,1}_{E}\Tr [P^{1} - \Sigma_{y}^{1}S^{\dagger}_{E}\Sigma^{1}_{y}S_{E}]& \nonumber \\
+f^{+,1}_{E}\Tr [P^{1}S^{\dagger}_{E}\Sigma^{1}_{y}S_{E}]& \nonumber \\
+f^{-,2}_{E}\Tr [\Sigma^{2}_{y}S^{\dagger}_{E}\Sigma^{1}_{y}S_{E}]& \nonumber \\
+f^{+,2}_{E}\Tr [P^{2}S^{\dagger}_{E}\Sigma^{1}_{y}S_{E}]&\!\!\!\big\}.
\end{eqnarray}
We can introduce into this a general S-matrix of the form
\begin{equation}
S_{E} = \left(
\begin{array}{cc}
\mathbb{r}^{1}_{E} & \mathbb{t}^{2}_{E} \\
\mathbb{t}^{1}_{E} & \mathbb{r}^{2}_{E}
  \end{array}\right)\,,
\end{equation}
in which unitarity restricts $\mathbb{r}^{\alpha\dagger}_{E}\mathbb{r}^{\alpha}_{E} + \mathbb{t}^{\alpha\dagger}_{E}\mathbb{t}^{\alpha}_{E} = \mathbb{1}$.  The average current now becomes
\begin{eqnarray}
I^{1} &=& \frac{ev_{F}}{4L}\sum_{E}f^{-,1}_{E}\big\{2N^{1} - \Tr [\Sigma_{y}\mathbb{r}^{1\dagger}_{E}\Sigma_{y}\mathbb{r}^{1}_{E}]\big\} \nonumber \\
&+& f^{+,1}_{E}\Tr [\mathbb{r}^{1\dagger}_{E}\Sigma_{y}\mathbb{r}^{1}_{E}] + f^{-,2}_{E}\Tr [\Sigma_{y}\mathbb{t}^{2\dagger}_{E}\Sigma_{y}\mathbb{t}^{2}_{E}]\nonumber \\
&+& f^{+,2}_{E}\Tr [\mathbb{t}^{2\dagger}_{E}\Sigma_{y}\mathbb{t}^{2}_{E}]\,,
\end{eqnarray}
where we have used $\Sigma_{y}$ to represent any square matrix of the form $\mathbb{1}\otimes\sigma_{y}$ and $\Sigma_{y}^{\alpha}$ to represent any projection of a matrix of that form.  This means that our $\Sigma_{y}$ matrices will possibly be different sizes; nevertheless they are carefully ordered in such a manner that all matrix products are still valid.

In our Majorana basis, only a single channel may be transmitting, which restricts the form of our $S$ matrix
\begin{eqnarray}
\mathbb{t}^{\alpha}_{E}&=&t^{\alpha}_{E}\mathbb{e}^{11}\,, \\
\mathbb{r}^{\alpha}_{E}&=&r^{\alpha}_{E}\mathbb{e}^{11} + \mathbb{\tilde{r}}^{\alpha}_{E}\,.
  \label{Smatrix1}
\end{eqnarray}
where $r^{\alpha}_{E}$ and $t^{\alpha}_{E}$ are scattering coefficients for the single transmitting channel and the elements of the matrix $\mathbb{e}^{ab}_{ij}=\delta_{ai}\delta_{bj}$.  The scattering matrices for the remaining channels $\mathbb{\tilde{r}}^{1}_{E}\neq\mathbb{\tilde{r}}^{2}_{E}$ are in general unequal as the leads differ in channel number and structure and are otherwise unrelated by additional symmetries.  These matrices have the structure
\begin{equation}
\mathbb{\tilde{r}}^{\alpha}_{E} = \left(
\begin{array}{cc}
0 & \vec{0}^{T} \\
\vec{0} & \mathbb{\tilde{R}}^{\alpha}_{E}
  \end{array}\right)\,,
  \label{Smatrix2}
\end{equation}
where $\vec{0}$ is the zero column vector and $\mathbb{\tilde{R}}^{\alpha}_{E}$ is an undetermined $(2N^{\alpha}-1)\times(2N^{\alpha}-1)$ dimensional reflection matrix.  $\mathbb{\tilde{R}}^{\alpha}_{E}$, as discussed in~\ref{sec:1dmodel}, is unconstrained by time-reversal symmetry for $\varphi\neq\pi$.  Additionally, due to the presence of some possible disorder and the underlying unusual ring geometry, the reflection matrices are also in the most general case unconstrained by spatial symmetries.  The presence of disorder may also reduce the number of channels which participate in interchannel scattering, effectively sending $N^{\alpha} \rightarrow N^{\alpha}_{open}$, where $N^{\alpha}_{open}$ is the number of transmitting electron channels through the disorder on the TI surface between the leads and the junction~\cite{imry86}.  Noting that $\mathbb{t}^{\alpha\dagger}_{E}\Sigma_{y}\mathbb{t}^{\alpha}_{E}=\mathbb{0}$ and $\Sigma_{y}\mathbb{e}^{11}\Sigma_{y}=\mathbb{e}^{22}$ while taking $\sum_{E}=L/(2\pi\hbar v_F)\int_{-\infty}^{+\infty} dE$, 
we arrive at a final answer for the total average current at the first lead
\begin{eqnarray}
I^{1} = \frac{e}{4h}\int_{-\infty}^{+\infty} dE\ f^{-,1}_{E}\big\{2N^{1} - 2\Real [r^{1*}_{E}\mathbb{\tilde{r}}^{1}_{22,E}] \nonumber \\
-\Tr [\Sigma_{y}\mathbb{\tilde{r}}^{1,\dagger}_{E}\Sigma_{y}\mathbb{\tilde{r}}^{1}_{E}]\big\}
\end{eqnarray}
of which the part containing singular Majorana behavior as $\varphi$ goes through $\pi$ is
\begin{equation}
I^{1}_{\gamma} = -\frac{e}{2h}\int_{-\infty}^{+\infty} dE\ f^{-,1}_{E}\big\{\Real [r^{1*}_{E}\mathbb{\tilde{r}}^{1}_{22,E}]\big\}.
\end{equation}

If we take the single-channel limit, {\em i.e.} the case of a quantum point contact, $N^{1}=N^{2}=1$ and $\mathbb{\tilde{r}}_{E}^{\alpha}$ is reduced to $\mathbb{\tilde{r}}^{1}_{22,E}=-\mathbb{\tilde{r}}^{2}_{22,E}=1$ and the average current becomes
\begin{equation}
I^{1} = \frac{e}{2h}\int_{-\infty}^{+\infty} dE\ f^{-,1}_{E}\bigg\{1 - \Real [r^{1*}_{E}]\bigg\}\,.
\end{equation}

In the more realistic many-channel limit, the elements of $\tilde{\mathbb{r}}^{\alpha}_{E}\sim O(1/\sqrt{N^{\alpha}})$ but with random phases.  This means that they will, in general, not add coherently such that in the large $N^{\alpha}$ limit, terms that contain $\tilde{\mathbb{r}}^{\alpha}_{E}$ will die off.  In this limit the average current reads
\begin{equation}
I^{1}_{many} = \frac{e}{2h}\int_{-\infty}^{+\infty} dE\ f^{-,1}_{E}N^{1}\,,
\end{equation}
and the $\varphi$ dependence is suppressed.

\section{Cross Noise Calculation}
\label{appendix:crossnoise}

In this section, we calculate in terms of our previous matrices the general expression for noise power, specializing at the end to the case of cross noise.  Zero-frequency noise power, $P^{\alpha\beta}=\int_{-\infty}^{+\infty}dt\big(\langle\hat{I}^{\alpha}(t)\hat{I}^{\beta}(0)\rangle - {I}^{\alpha}{I}^{\beta}\big)$, can be calculated using Wick's theorem as follows:
\begin{eqnarray}
P^{\alpha\beta} &=& \frac{e^{2}v_{F}}{L} \sum_{E,E'} \langle\gamma^{a}_{-E}\gamma^{b}_{E}\gamma^{c}_{-E'}\gamma^{d}_{E'}\rangle A^{ab,\alpha}_{E}A^{cd,\beta}_{E'} \nonumber\\
&=& \frac{e^{2}v_{F}}{L}\sum_{E}Q^{ad}_{E}Q^{bc}_{-E}A^{ab,\alpha}_{E}A^{cd\beta}_{E} - Q^{ac}_{E}Q^{bd}_{-E}A^{ab,\alpha}_{E}A^{cd,\beta}_{-E}\nonumber\\
&=& 2{e^2 v_F\over L}\sum_{E}\Tr [A^{\alpha}_{E}Q_{-E}A^{\beta}_{E}Q^{T}_{E}].
\label{matrixyNoise}
\end{eqnarray}

From here, we can specialize to the $\alpha\neq\beta$ case and calculate the cross noise power.  For $\alpha=1$, $\beta=2$,
\begin{equation}
P^{12}=4{e^2 v_F\over L} \sum_{E}\Tr [S^{\dagger}_{E}\Sigma^{1}_{y}S_{E}Q^{1}_{-E}S^{\dagger}_{E}\Sigma^{2}_{y}S_{E}Q_{E}^{2,T}]\,.
\end{equation}
After specializing the S-matrix elements and taking $\sum_{E}=(L/(2\pi\hbar v_F))\int dE$, $P^{12}$ becomes
\begin{equation}
P^{12}_{\gamma} = -\frac{e^{2}}{4h}\int_{-\infty}^{+\infty} dE\ f^{-,1}_{E}f^{-,2}_{E}t^{1}_{E}t^{2}_{E}\mathbb{\tilde{r}}^{1*}_{22,E}\mathbb{\tilde{r}}^{2*}_{22,E}\,,
\end{equation}
where we have used $P^{12}_{\gamma}$ to note that {\it all} of $P^{12}$ behaves singularly as the system goes through its critical point.  

In the single-channel limit discussed in Appendix~\ref{appendix:averagecurrent},
\begin{equation}
P^{12} = \frac{e^{2}}{4h}\int_{-\infty}^{+\infty} dE\ f^{-,1}_{E}f^{-,2}_{E}t^{1}_{E}t^{2}_{E}.
\end{equation}
In the many-channel limit, $P^{12}\rightarrow 0$ as $O(1 / \sqrt{N^{1}N^{2}})$.

\section{Diagonal Noise Calculation}
\label{appendix:diagonalnoise}

The diagonal noise calculation follows very closely the cross noise calculation.  However, unlike $P^{12}$, $P^{11}$ does not have a simple general form.  We begin with Eq.~(\ref{matrixyNoise})
\begin{equation}
P^{11} = 2{e^2 v_F\over L}\sum_{E}\Tr [A^{1}_{E}Q_{-E}A^{1}_{E}Q^{T}_{E}],
\end{equation}
which given Eq.~(\ref{Smatrix1}) and~(\ref{Smatrix2}) becomes
\begin{eqnarray}
P^{11} &=& {e^2\over 8h}\int_{-\infty}^{+\infty} dE\nonumber\\
&\bigg\{& f^{+,1}_{-E}f^{+,1}_{E}\big\{4N^{1} -2\Tr [\Sigma^{1}_{y}\mathbb{r}^{1\dagger}_{E}\Sigma_{y}\mathbb{r}^{1}_{E}] -2|t_{E}|^2\big\} \nonumber \\
&+& f^{-,1}_{E}f^{-,1}_{E}\big\{-2N^{1} + 2\Tr [\Sigma_{y}\mathbb{r}^{1\dagger}_{E}\Sigma_{y}\mathbb{r}^{1}_{E}] \nonumber \\
&-&\Tr [\Sigma_{y}\mathbb{r}^{1\dagger}_{E}\Sigma_{y}\mathbb{r}^{1}_{E}\Sigma_{y}\mathbb{r}^{1\dagger}_{E}\Sigma_{y}\mathbb{r}^{1}_{E}]\big\} \nonumber \\
&+&(f^{-,1}_{E}f^{+,2}_{E} - f^{-,1}_{E}f^{+,1}_{E})\big\{2\Tr [\Sigma_{y}\mathbb{t}^{1}_{E}\mathbb{t}^{1\dagger}_{E}\Sigma_{y}\mathbb{r}^{1\dagger}_{E}\Sigma_{y}\mathbb{r}^{1}_{E}]\big\} \nonumber \\
&+&f^{+,1}_{-E}f^{+,2}_{E}2|t_{E}|^{2}\bigg\}
\end{eqnarray}
under the usual substitution for $\sum_{E}$.  Further simplifying this, we can obtain expressions for the traces of scattering matrices:
\begin{eqnarray}
\Tr [\Sigma_{y}\mathbb{t}^{1}_{E}\mathbb{t}^{1\dagger}_{E}\Sigma_{y}\mathbb{r}^{1\dagger}_{E}\Sigma_{y}\mathbb{r}^{1}_{E}] = |t_{E}|^{2}(\mathbb{\tilde{r}}^{1}_{E}\Sigma_{y}\mathbb{\tilde{r}}^{1\dagger}_{E})_{22} \nonumber \\
\nonumber\\
\Tr [\Sigma_{y}\mathbb{r}^{1\dagger}_{E}\Sigma_{y}\mathbb{r}^{1}_{E}] =2\Real [r^{1*}_{E}\mathbb{\tilde{r}}^{1}_{22,E}] 
+ \Tr[\Sigma_{y}\mathbb{\tilde{r}}_{E}^{1}\Sigma_{y}\mathbb{\tilde{r}}_{E}^{1\dagger}]  \nonumber \\
\nonumber\\
\Tr [\Sigma_{y}\mathbb{r}^{1\dagger}_{E}\Sigma_{y}\mathbb{r}^{1}_{E}\Sigma_{y}\mathbb{r}^{1\dagger}_{E}\Sigma_{y}\mathbb{r}^{1}_{E}] = \nonumber \\
2\Real\bigg[(r^{1*}_{E}\mathbb{\tilde{r}}^{1}_{22,E})^{2} + r^{1}_{E}(\mathbb{\tilde{r}}^{1}_{E}\Sigma_{y}\mathbb{\tilde{r}}^{1\dagger}_{E}\Sigma_{y}\mathbb{\tilde{r}}^{1}_{E})_{22}\bigg] \nonumber \\
\Tr [\Sigma_{y}\mathbb{\tilde{r}}_{E}^{1}\Sigma_{y}\mathbb{\tilde{r}}_{E}^{1\dagger}\Sigma_{y}\mathbb{\tilde{r}}_{E}^{1}\Sigma_{y}\mathbb{\tilde{r}}_{E}^{1\dagger}].
\end{eqnarray}

This very complicated expression does not have, like the current and cross noise, clearly separable singular pieces in its general N-channel form.  However in the single-channel limit as described in Appendix~\ref{appendix:averagecurrent}, it simplifies significantly:
\begin{eqnarray}
P_{11} &=& {e^2\over{4h}} \int_{-\infty}^{+\infty} dE  \big[|1-r^{1}_E|^2 f^{+,1}_{-E}f^{+,1}_{E} \nonumber\\
&-&(1-r^{1}_E)^2 f^{-,1}_{E}f^{-,1}_{E} + |t_E|^2 f^{+,1}_{-E}f^{+,2}_{E}\big]\,, 
\end{eqnarray}
where we have exploited in the second term that $f^{-,\alpha}_{-E}=f^{-,\alpha}_{E}$ and that $r^{\alpha}_{-E}=r^{\alpha *}_{E}$.  In the many-channel limit, the diagonal noise becomes quite simple:
\begin{eqnarray}
P^{11}_{many} = {e^2\over 4h}\int_{-\infty}^{+\infty} dE\ f^{+,1}_{-E}f^{+,1}_{E}\big\{2N^{1} -|t_{E}|^2\big\}& \nonumber \\
+f^{+,1}_{-E}f^{+,2}_{E}|t_{E}|^{2}&\,,
\end{eqnarray}
where unlike with $P^{12}$, the $\varphi$ dependence is mostly preserved and is much more clearly extracted than in the general, N-channel case.  Additionally, the singular part of the diagonal noise, $P^{11}_{\gamma}$, goes to zero in the many-channel limit if $V^{1}=V^{2}$.  

\end{appendix}


\begin{thebibliography}{10}
\bibitem{rashbanote} A similar setup may be possible using a 2D Rashba semiconductor proximity effect device\cite{sau10,alicea10}, though in that case the magnetic field will affect both the flux through the ring and the topological superconductivity.
\bibitem{similar model}  A similar model, describing the transition between a 1D trivial and topological superconductor, has been analyzed in Ref. \onlinecite{akhmerov11}.
\end{thebibliography}
\end{document}